# Characteristic length for pinning force density in Nb3Sn

E. F. Talantsev[1,2], E. G. Valova-Zaharevskaya[1], I. L. Deryagina[1], E. N. Popova[1]

[1] M.N. Miheev Institute of Metal Physics, Ural Branch, Russian Academy of Sciences,
18, S. Kovalevskaya St., Ekaterinburg, 620108, Russia

[2]NANOTECH Centre, Ural Federal University, 19 Mira St., Ekaterinburg, 620002, Russia

**Abstract**

The pinning force density $\vec{F_p}(J_c, B) = \vec{J_c} \times \vec{B}$ (where $J_c$ is the critical current density and $B$ is the applied magnetic field) is one of the main parameters that characterize the resilience of a superconductor to carry a dissipative-free transport current in an applied magnetic field. Kramer (1973 *J. Appl. Phys.* **44** 1360), and Dew-Hughes (1974 *Phil. Mag.* **30** 293) proposed a widely used scaling law for the pinning force density amplitude: $|\vec{F_p}(B)| = F_{p,max} \times \frac{(p+q)^{p+q}}{p^p q^q} \times \left(\frac{B}{B_{c2}}\right)^p \times \left(1 - \frac{B}{B_{c2}}\right)^q$, where $F_{p,max}$, $B_{c2}$, $p$, and $q$ are free-fitting parameters. Since the late 1970-s till now, several research groups have reported experimental data on the dependence of $F_{p,max}$ on the average grain size, $d$, in Nb3Sn-based conductors. Godeke (2006 *Supercond. Sci. Techn.* **19** R68) proposed that the dependence obeys the law $|F_{p,max}(d)| = A \times ln(1/d) + B$. However, this scaling law has several problems, for instance, the logarithm is taken from a non-dimensionless variable, and $|F_{p,max}(d)| < 0$ for large grain sizes, and $|F_{p,max}(d)| \to \infty$ for $d \to 0$. Here, we reanalysed the full inventory of publicly available $F_{p,max}(d)$ data for Nb3Sn conductors and found that the dependence can be described by $F_{p,max}(d) = F_{p,max}(0) \times exp(-d/\delta)$ law, where the characteristic length, $\delta$, varies within a remarkably narrow range, that is, $\delta = 175 \pm 13\ nm$, for samples fabricated by different technologies. The interpretation of the result is based on the idea that the in-field supercurrent flows within a thin surface layer (thickness of $\delta$) near the grain boundary surfaces (similar to London's law, where the self-field supercurrent flows within a thin surface layer with a thickness of the London penetration depth, $\lambda$, and the surface is a superconductor-vacuum surface). An alternative interpretation is that $\delta$ represents the characteristic length of the exponential decay flux pinning potential from the dominant defects in Nb3Sn superconductors, which are grain boundaries.

# Characteristic length for pinning force density in Nb₃Sn

**1. Introduction**

Multifilamentary superconducting Nb$_3$Sn-based wires are used in many high-energy physics and fusion energy projects, including international mega-science projects such as the Large Hadron Collider (LHC) [1] and International Thermonuclear Experimental Reactor (ITER) [2]. The Nb$_3$Sn-based superconductors should, first of all, have high current-carrying capacity in high magnetic fields. In particular, the modernization of LHC [3] involves the replacement of a part of NbTi conductors with Nb$_3$Sn-based conductors. Particularly, to create high-field large-aperture quadrupole MQXF [4] and high-field 11-T dipoles [5] for the high-luminosity LHC Upgrade Project, the development of a new generation of high-field Nb$_3$Sn-based superconductors is required for the effective use of the advantages of Nb$_3$Sn wires over previously used NbTi, to provide a minimum critical current of approximately 360 A and higher in a field of 15 T at 4.2 K [3]. The critical current density $J_c$ of the modern designed Nb$_3$Sn strand has achieved record values of non-Cu $J_c(B = 12\text{ T}, T = 4.2\text{ K}) = 3000 \text{ A/mm}^2$ and $J_c(B = 15\text{ T}, T = 4.2\text{ K}) = 1700 \text{ A/mm}^2$ [6]. However, to create a Future Circular Collider (FCC) at CERN, Nb$_3$Sn-based wires with $J_c(B = 16\text{ T}, T = 4.2\text{ K}) = 1500 \text{ A/mm}^2$ or $J_c(B = 12\text{ T}, T = 4.2\text{ K}) = 3500 \text{ A/mm}^2$ are required [7].

For thermonuclear power engineering, bronze-processed Nb$_3$Sn-based wires were developed for superconducting magnets of the ITER project, providing a J$_c$ of approximately 750 A/mm² and higher in a field of 12 T [2]. The next mega-science project after the ITER should be the DEMO experimental facility, the primary goal of which is to demonstrate the possibility of obtaining a positive power balance from a thermonuclear reactor as the whole system. This goal requires the development of superconducting Nb$_3$Sn-based conductors with even better characteristics [8].

Extensive (nearly five decades) R&D studies of Nb$_3$Sn-based conductors have shown that the key factors affecting the in-field critical current in these wires are the local composition, structure, and morphology of the superconducting A-15 phase [9–17].

These studies also showed that at high magnetic fields, the main pinning centers in $Nb_3Sn$-based composites are grain boundaries, and the conventional approach to increasing $J_c(B,T)$, in $Nb_3Sn$ is to increase the density of grain boundaries, that is, to ensure grain refinement. To achieve this, various manufacturing methods and designs of multifilamentary wires have been proposed [7], targeting the creation of small average grain sizes in the superconducting phase [18–21].

Superconducting wires based on $Nb_3Sn$ are produced by one of the following methods: bronze route, internal tin (IT), and power in tube (PIT) [22–24]. In the bronze route, an initial billet formed of Nb, Nb-Ti or Nb-Ta rods assembled in a bronze Cu-Sn matrix and external copper tube is extruded and drawn to a small diameter. The $Nb_3Sn$ phase is formed by Sn diffusion from the matrix to Nb filaments under heat treatment (HT), which is usually referred to as diffusion annealing. The solid-state diffusion of Sn at relatively low temperatures of HT prevents excessive grain growth and increases the pinning efficiency. The main disadvantage of the bronze method is the limited solubility of Sn in the bronze matrix when the Sn concentration increases to more than 8 mass. %, brittle phases are precipitated, which impedes plastic deformation and leads to cracking of the composite wire at the manufacturing stage. Therefore, to ensure a sufficient amount of Sn for the formation of the $Nb_3Sn$ phase, the ratio of the volume fractions of bronze and niobium should not be less than 3:1. Owing to these restrictions, bronze-processed wires have lower $J_c$-values than those potentially possible for the $Nb_3Sn$ phase. An important step in the development of bronze technology was the development of the Osprey method for producing high-tin bronze, which retains its plasticity up to 15−17 mas. % Sn. Using such bronze makes it possible to increase the number of Nb filaments in the strand, provide a complete transformation of Nb filaments into the superconducting phase, and increase the Sn concentration in the $Nb_3Sn$ layers, which results in an increase of $J_c$ [25]. However, even in the $Nb_3Sn$ strands fabricated using a high-Sn bronze matrix, it is not possible to avoid large $Nb_3Sn$ composition gradients across the superconducting layer. These gradients, in turn, produce large gradients in the superconducting properties that limit the overall current density, particularly in high fields [9]. The deficiency of tin leads to the formation of a relatively large fraction of non-stoichiometric Nb$_3$Sn compounds [26], which are stable from 18 to 25 at. % Sn, and the low-tin part of superconducting layers loses its superconductivity in high fields [27].

The IT process was developed to avoid frequent in-process annealing during wire drawing and to enhance the available Sn concentration with respect to the bronze process using separate Sn, Cu, and Nb billet stacking elements rather than specially melted high-Sn bronze matrix alloys [28]. The modified design of modern IT strands (e.g. strands with distributed diffusion barriers) makes it possible to obtain $J_c$ beyond 2200 A/mm$^2$ and achieve a record-braking value of 3000 A/mm$^2$ (non-copper, l2 T, 4.2 K) [12,29]. The highest critical current density strands have Nb$_3$Sn layers with minimal chemical and microstructural inhomogeneity and a high fraction of the close to stoichiometric phase.

To increase the $J_c$ of superconductors designed to operate in high magnetic fields (15-16 T and higher), new designs of superconducting strands are created based on the IT technology, which are referred to as high-$J_c$ strands. According to Ref. [30], the OST company produces high-$J_c$ strands using the so called Restacked Rod Process (RRP) design of the wires. In the RRP strands, the many Cu-clad niobium filaments surrounding the tin source inside the subelement grow through the inter-filamentary Cu and formed a single Nb$_3$Sn tube in the volume of the strand. Each subelement was surrounded by a Nb-Ta diffusion barrier, which was designed to partially react, and $J_c$ values of these strands were approximately 3000 A/mm$^2$. The compositional analysis of the high-current wires indicated that the Sn content was relatively uniform at approximately 24 ± 1 at. % Sn in the A15 volume [31].

The PIT process [32] combines an abundant Sn source with a relatively high current density (over 2500 A/mm$^2$) and fine filaments (approximately 35 μm). The abundant Sn source results in a relatively high Sn content in the A15 phase. This indicates that the PIT wires contained a relatively large A15 fraction rich in Sn. The maximum non-Cu $J_c$ is from 2600 A/mm$^2$ (at 12 T, 4.2 K) in 1.25 mm wires, for superconducting wires, which were developed for the Next European

Dipole (NED) program. The main advantages of the PIT process are shorter heat treatments because of the close location of the Sn source to the niobium, no pre-heating treatment is required compared to other methods, and relatively small filaments (30–50 μm) can be obtained, which leads to low hysteresis losses. The main disadvantage of the PIT manufacturing routine is its higher cost compared with other fabrication technologies [33,34].

The resilience of any superconducting wire to carry a dissipative-free transport current at an applied magnetic field can be quantified by the pinning force density, $\vec{F_p}$, (defined as a vector product of the transport critical current density, $\vec{J_c}$, and the applied magnetic field, $\vec{B}$):

$$\vec{F_p}(J_c, B) = \vec{J_c} \otimes \vec{B}. \qquad (1)$$

For an isotropic superconductor and maximal Lorentz force geometry, i.e. when $\vec{J_c} \perp \vec{B}$, Kramer [35] and Dew-Hughes [36] proposed a widely used scaling expression for the amplitude of the pining force density [37]:

$$\left|\vec{F_p}(B)\right| = F_{p,max} \times \frac{(p+q)^{p+q}}{p^p q^q} \times \left(\frac{B}{B_{c2}}\right)^p \times \left(1 - \frac{B}{B_{c2}}\right)^q, \qquad (2)$$

where $F_{p,max}, B_{c2}, p$, and $q$ are free-fitting parameters, and $B_{c2}$ is the upper critical field, and $F_{p,max}$ is pinning force density amplitude.

Figure 1 shows a typical $\left|\vec{F_p}(B, 4.2\ K)\right|$ for Nb$_3$Sn superconductors reported by Flükiger *et al* [38], where the data fit to Equation (2) and deduced free-fitting parameters, $F_{p,max}, B_{c2}, p$, and $q$ are shown.

While the upper critical field, $B_{c2}$, is one of the fundamental parameters for a given superconducting phase, three other parameters in Equation (2), that is, $F_{p,max}, p$, and $q$, depend on the superconductor microstructure, presence of secondary phases, and so on. In accordance with the approach proposed by Dew-Hughes [36], the shape of the $\left|\vec{F_p}(B)\right|$ (defined by $p$ and $q$) reflects the primary pinning mechanism in a sample.

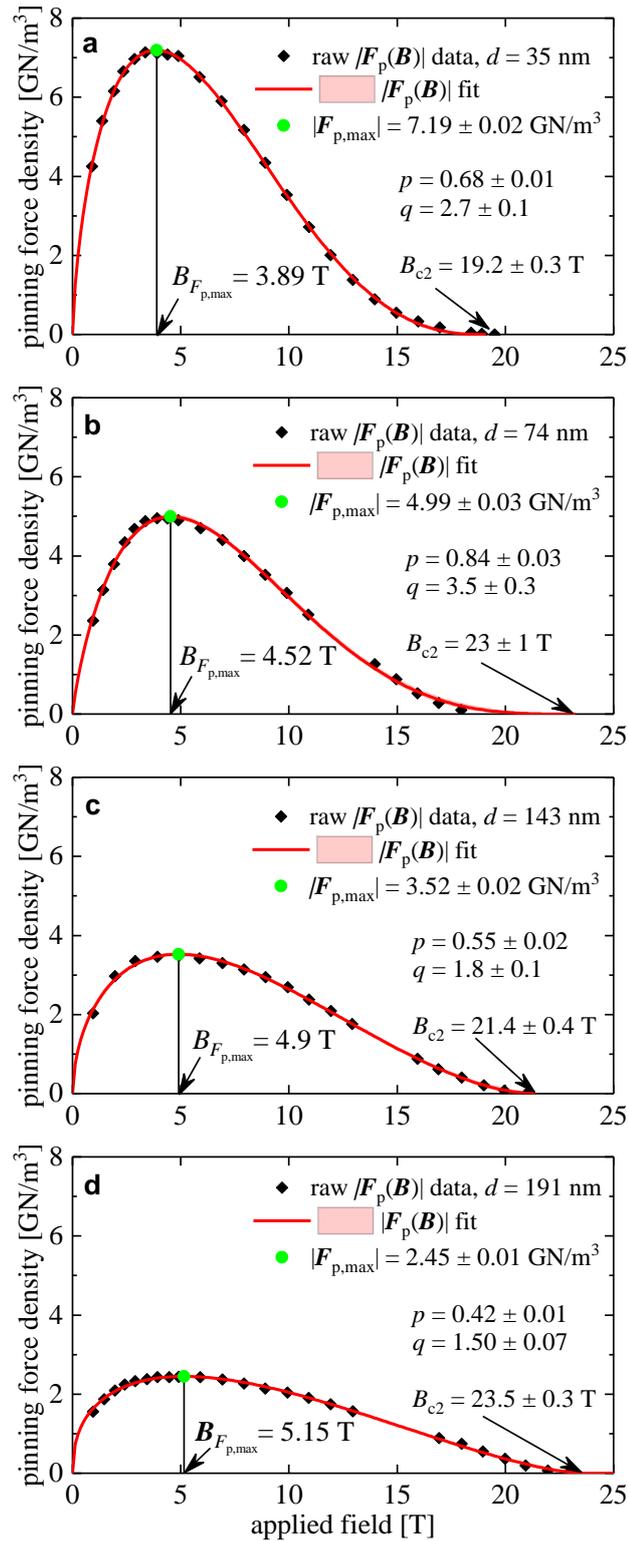

**Figure 1.** Pinning force density $F_p$ versus $B$ for bronze-route wires of different average grain sizes, $d$: **(a)** $d$ = 35 nm; deduced $F_{p,max}$ = 7.19 ± 0.02 GN/m$^3$, $B_{c2}$ = 19.2 ± 0.3 T, $p$ = 0.68 ± 0.01, $q$ = 2.7 ± 0.1; fit quality is 0.9997; **(b)** $d$ = 74 nm; deduced $F_{p,max}$ = 4.99 ± 0.03 GN/m$^3$, $B_{c2}$ = 23 ± 1 T, $p$ = 0.84 ± 0.03, $q$ = 3.5 ± 0.3; fit quality is 0.9982; **(c)** $d$ = 143 nm; deduced $F_{p,max}$ = 3.52 ± 0.02 GN/m$^3$, $B_{c2}$ = 21.4 ± 0.4 T, $p$ = 0.55 ± 0.02, $q$ = 1.8 ± 0.1; fit quality is 0.9987; **(d)** $d$ = 191 nm; deduced $F_{p,max}$ = 2.45 ± 0.01 GN/m$^3$, $B_{c2}$ = 23.5 ± 0.3 T, $p$ = 0.42 ± 0.01, $q$ = 1.50 ± 0.07; fit quality is 0.9986. The $p$ and $q$ parameters for the fit were determined using the Kramer-Dew-Hughes equation (Equation (2)). Raw data reported by Flükiger *et al* [38]. 95% confidence bands are shown by pink shadow areas.

While the upper critical field, $B_{c2}$, is one of the fundamental parameters for a given superconducting phase, three other parameters in Equation (2), that is, $F_{p,max}$, $p$, and $q$, depend on the superconductor microstructure, presence of secondary phases, and so on. In accordance with the approach proposed by Dew-Hughes [36], the shape of the $|\vec{F_p}(B)|$ (defined by $p$ and $q$) reflects the primary pinning mechanism in a sample. Dew-Hughes [36] calculated theoretical characteristic values for $p$ and $q$ for different pinning mechanisms, in particularly for point defect (PD) and grain boundary (GB) pinning.

The evolution of the dominant pinning mechanism from GB- to PD-pinning in $Nb_3Sn$ under neutron irradiation was recently reported by Wheatley *et al* [39], who showed that the unirradiated $Nb_3Sn$ alloy exhibits $|\vec{F_p}(B,T)|$ form indicating the dominance of the GB-pinning, and after the neutron irradiation the $|\vec{F_p}(B,T)|$ form transforms towards the PD-pinning mode.

The fourth parameter in Equation (2), which is the $F_{p,max}$, represents the maximal performance of a given superconductor in an applied magnetic field. It is well-established experimental fact [38,40–46] that the $F_{p,max}$ in $Nb_3Sn$ depends on the average grain size, $d$, of the material. The traditional approach to representing the $F_{p,max}$ vs. $d$ dependence is to use a reciprocal semi-logarithmic plot (Figure 2). Godeke [41] proposed the following form for the $F_{p,max}$ vs. $d$ dependence:

$$F_{p,max}(d) = A \times ln(1/d) + B, \qquad (3)$$

where free-fitting parameter $A = 22.7$ and $B = -10$.

Following traditional methodology [37], Godeke [41] proposed that because grain boundaries are primary pinning centers in $Nb_3Sn$, there is an optimum grain size, $d_{opt}$, at which the maximum performance for a given wire can be achieved for a given applied magnetic field, $B$. This field [41] is equal to the flux line spacing in the hexagonal vortex lattice, $a_{hexagonal}$ [47], at the applied field $B$, which can be designated as the matching field, $B_{match}$, at the maximum pinning force density:

$$d_{opt} = a_{hexagonal} = \left(\frac{4}{3}\right)^{1/4} \times \left(\frac{\phi_0}{B_{match}}\right)^{1/2}, \quad (4)$$

where $\phi_0 = \frac{h}{2e}$ is superconducting flux quantum.

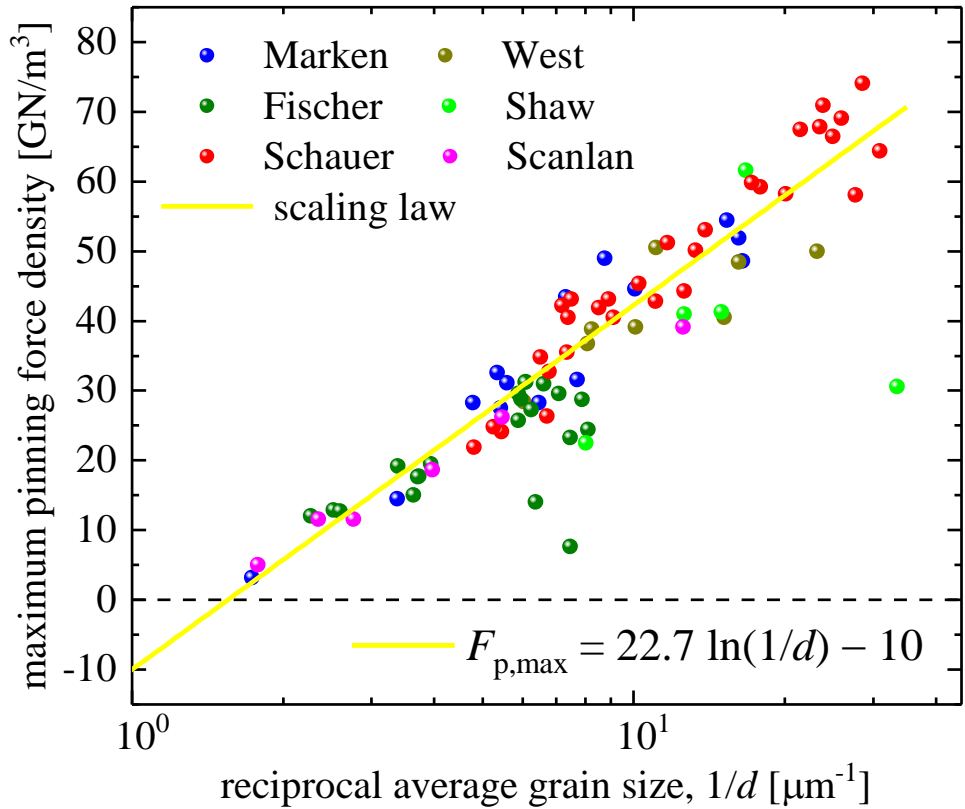

**Figure 2.** Maximum pinning force density, $F_{p,max}$, vs. reciprocal average grain size, $1/d$, for datasets reported by Marken [42], West *et al* [43], Fischer [40], Shaw [44], Schauer *et al* [45], and Scanlan *et al* [46]. Fitting curve (Equation (3)) was proposed by Godeke [41], who also presented full dataset in the log-linear plot.

Here, we show that neither Equation (3) nor Equation (4) provides a valuable description of the available experimental $F_{p,max}(d)$ data measured over several decades in Nb$_3$Sn conductors. We also propose a new model to describe a full set of publicly available experimental datasets on the maximum pinning force density vs. grain size, $F_{p,max}(d)$.

## 2. Problems associated with current models

Equation (4) implies that if the grain size, $d_{opt}$, in some Nb$_3$Sn conductors has been determined, then the matching applied magnetic field, $B_{match}$, can be calculated as:

$$B_{match}(d_{opt}) = \left(\frac{4}{3}\right)^{1/2} \times \left(\frac{\phi_0}{d_{opt}^2}\right). \tag{5}$$

Following this logic [41], one can expect that the maximal performance in magnetic flux pinning, that is, $F_{p,max}$, should be observed at $B_{match}$:

$$B_{match}(d_{opt}) = B_{F_{p.max}}(d_{opt}) = \left(\frac{4}{3}\right)^{1/2} \times \left(\frac{\phi_0}{d_{opt}^2}\right). \tag{6}$$

In Figure 1, we fitted $|\vec{F_p}(B)|$ data [38] to Equation (1) for Nb$_3$Sn conductors with different grain sizes, $d$, from which the $B_{F_{p.max},exp}(d)$ were extracted. In Figure 3, we show $B_{F_{p.max},exp}(d)$ and calculated $B_{F_{p.max},calc}(d)$ (Equation (6)), from which it can be concluded that the traditional understanding of the primary mechanism governing dissipative-free high-field current capacity in Nb$_3$Sn conductors [41] is incorrect.

The validity of the $F_{p,max}(d)$ scaling law proposed by Godeke (Equation (3) [41]) was analyzed and it was concluded that there are at least three fundamental problems with the law:

1. The logarithmic function used in Equation (3), as well as all other mathematical functions, can operate only with the dimensionless variable, whereas the variable in Equation (3) has the dimension of inverse length. For instance, the variable $B$ in the Kramer-Dew-Hughes scaling law (Equation (2)) has the dimension cancelation term $\frac{1}{B_{c2}}$. The same general approach can be found for all equations in Ginzburg-Landau [47], Bardeen-Cooper-Schrieffer [48], and other physical theories [49], all of which implement this general rule.

For instance, the lower critical field, $B_{c1}$, in superconductors has traditional form [50]:

$$B_{c1}(T) = \frac{\phi_0}{4\pi\lambda^2(T)} \times \left(\ln\left(\frac{\lambda(T)}{\xi(T)}\right) + \alpha\left(\frac{\lambda(T)}{\xi(T)}\right)\right), \tag{7}$$

where

$$\alpha(\kappa) = \alpha_\infty + e^{\left(-c_0 - c_1 \times ln\left(\frac{\lambda(T)}{\xi(T)}\right) - c_2 \times \left(ln\left(\frac{\lambda(T)}{\xi(T)}\right)\right)^2\right)} \pm \varepsilon, \tag{8}$$

where $\lambda(T)$ is the London penetration depth, $\xi(T)$ is the superconducting coherence length, $\alpha_\infty = 0.49693$, $c_0 = 0.41477$, $c_1 = 0.775$, $c_2 = 0.1303$, and $\varepsilon \leq 0.00076$. Equations (7), (8) were recently simplified to the following form [51]:

$$B_{c1}(T) = \frac{\phi_0}{4\pi\lambda^2(T)} \times \left(\ln\left(1 + \sqrt{2}\frac{\lambda(T)}{\xi(T)}\right)\right). \tag{9}$$

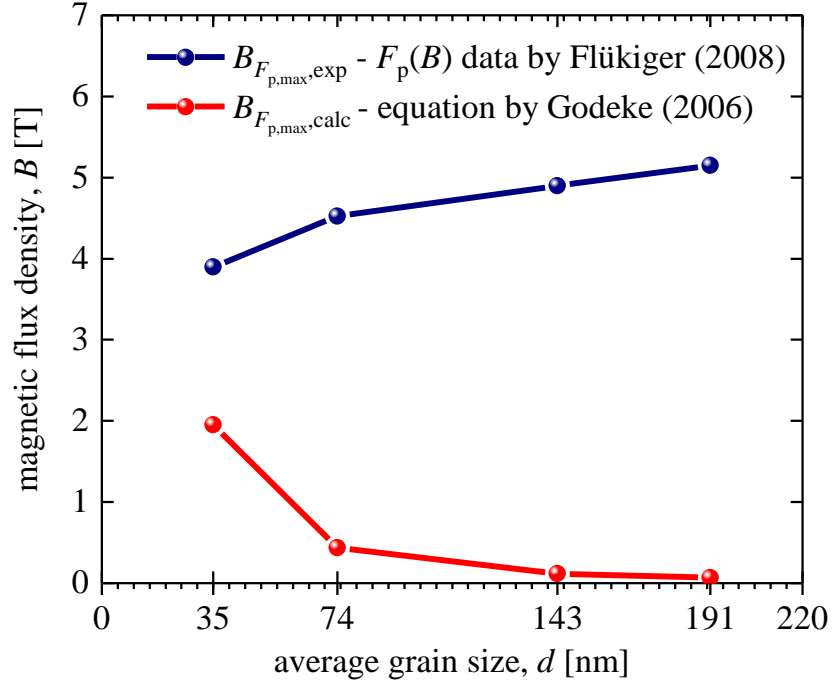

**Figure 3.** $B_{F_{p.max},calc}$ was calculated using Equation (4) (red) [41] and $B_{F_{p.max},exp}$ was extracted from experimental data reported by Flükiger [38] for Nb$_3$Sn conductors fabricated by bronze technology.

In Equations (7) and (9) the variable under the logarithm is dimensionless. The same can be found in the equation for the universal self-field critical current density, $J_c(sf,T)$, in thin film superconductors [52]:

$$J_c(sf,T) = \frac{\phi_0}{4\pi\mu_0\lambda^3(T)} \times \left(\ln\left(\frac{\lambda(T)}{\xi(T)}\right) + 0.5\right), \tag{10}$$

where $\mu_0$ is the permeability of the free space. It should be noted that Equation (10) was recently confirmed by Paturi and Huhtinen [53] for YBa$_2$Cu$_3$O$_{7-\delta}$ thin films that exhibit different mean-free paths for charge carriers. Equation (10) was also recently used by Troyan *et al* [54] to deduce the ground state London penetrations depth, $\lambda(0)$, and ground state superconducting energy gap, $\Delta(0)$, in highly-compressed hydrogen-rich superconductor $SnH_4$.

The same principle of unitless variable is implemented in all general physics laws, for instance, in Planck's law [49]:

$$B_\nu(\nu, T) = \frac{2h\nu^3}{c^2} \times \frac{1}{e^{\left(\frac{h\nu}{k_B T}\right)} - 1}, \tag{11}$$

where $B_\nu(\nu, T)$ is the spectral radiance of a body, $h$ is the Planck constant, $\nu$ is the frequency, $c$ is the speed of light in the medium, $k_B$ is the Boltzmann constant, and where the variable under the exponential function, $\frac{h\nu}{k_B T}$, is dimensionless.

Based on all above, Equation (3) has a fundamental mistake based on a simple fact that $ln(1/d)$ is an absurdum expression.

2. Even if the problem mentioned above (i.e. in #1) is omitted, there are two other problems associated with Equation (3). One problem is the limit of Equation (3) for a large grain size. In Figure 4, we replotted $|F_{p,max}(d)|$ data from Figure 2 in a linear-linear plot and showed both side extrapolations of Equation (3) within the range of $20\ nm \leq d \leq 800\ nm$, which is the usual range of grain sizes in Nb₃Sn conductors. In Figures 2, 4 one can see that:

$$\left|F_{p,max}(d)\right|_{d \geq 550\ nm} = (A \times ln(1/d) + B)|_{d \geq 550\ nm} < 0, \tag{12}$$

which is the absurdum. We also noted that the free-fitting parameters deduced by us ($A = 21.9 \pm 1.2$, $B = -9.9 \pm 2.7$) from the fit of the $|F_{p,max}(d)|$ dataset to Equation (3), are different from the values reported by Godeke [41], $A = 22.7$, $B = -10$, who analysed the same $|F_{p,max}(d)|$ dataset.

3. A similar validity problem of Equation (3) is for small grain sizes:

$$\lim_{d \to 0} |F_{p,max}(d)| = \lim_{d \to 0} (A \times ln(1/d) + B) = \infty, \tag{13}$$

which is unphysical because when $d$ becomes comparable to the double coherence length (which is the size of a normal vortex core):

$$d_{\min}(4.2\ K) \cong 2 \times \xi(T) = 2 \times \frac{\xi(0)}{\sqrt{1 - \frac{T}{T_c}}} = \frac{2 \times 3.0\ nm}{\sqrt{1 - \frac{4.2\ K}{18\ K}}} = 6.9\ nm, \tag{14}$$

where $\xi(0) = 3.0$ nm [55] and $T_c = 18$ K [55] were used, a further decrease in the grain size $d$ should not cause any changes in the magnetic flux pinning, and thus in $|F_{p,max}(d)|$ amplitude.

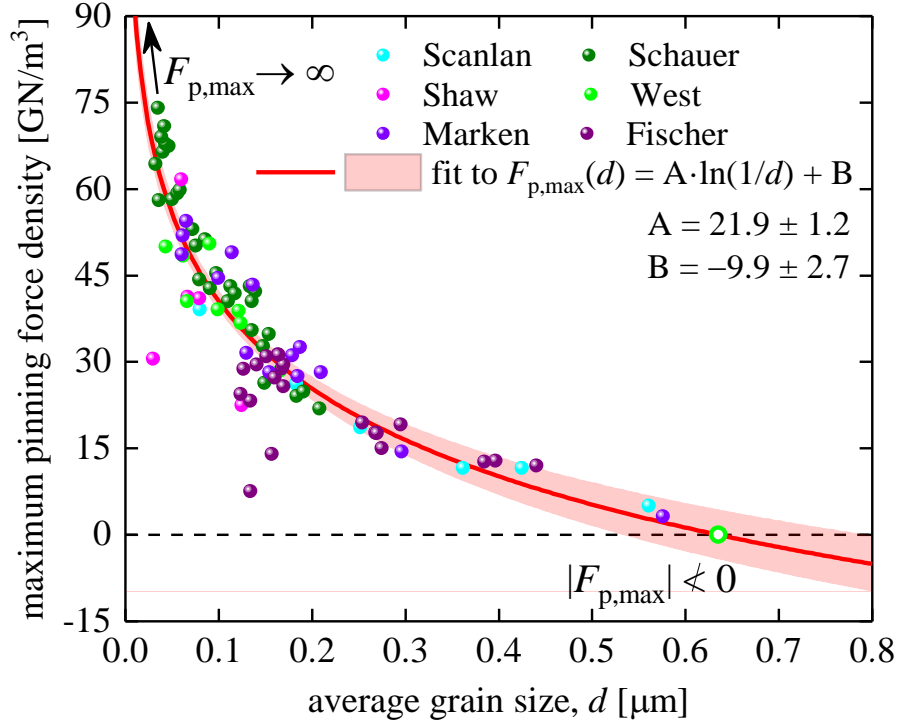

**Figure 4**. $|F_{p,max}(d)|$ data from Figure 2 (reported by Fischer [40] and Godeke [41]) in linear-linear plot, and the fitting curve to Equation (3) [41], where we also showed both side extrapolations within the average grain size range of $20\ nm \leq d \leq 800\ nm$ of $Nb_3Sn$. Raw data reported by Marken [42], West et al [43], Fischer [40], Shaw [44], Schauer et al [45], and Scanlan et al [46].

### 3. Results

By experimenting with many analytical functions that can approximate $|F_{p,max}(d)|$ dependence shown in Figures 2 and 4, we found a remarkably simple, robust, heuristic, and physically sounded expression:

$$|F_{p,max}(d)| = |F_{p,max}(0)| \times e^{-\frac{d}{\delta}} \qquad (15)$$

where $|F_{p,max}(0)|$ and $\delta$ are free fitting parameters. This function exhibits physically sounded limits:

$$\lim_{d \to \infty} |F_{p,max}(d)| = \lim_{d \to \infty} \left(|F_{p,max}(0)| \times e^{-\frac{d}{\delta}}\right) = 0, \qquad (16)$$

$$\lim_{d \to 0} |F_{p,max}(d)| = \lim_{d \to 0} \left(|F_{p,max}(0)| \times e^{-\frac{d}{\delta}}\right) = |F_{p,max}(0)| < \infty. \qquad (17)$$

We proposed interpretations for $|F_{p,max}(0)|$ and of $\delta$ parameters in the *Discussion* Section. Before that, in this Section we show the robustness of Equation (15) to fit publicly available datasets for Nb$_3$Sn conductors. Data fitting was performed in Origin2017 software.

### 3.1. Bronze technology samples

Bronze technology for Nb$_3$Sn-based wires has been described in detail elsewhere [1]. For our analysis, we used $|F_{p,max}(d)|$ dataset reported by Godeke [41]. Godeke [56] pointed out that Fischer [40] collected raw $|F_{p,max}(d)|$ data (shown in Figures 2 and 4), and these data are "*all pre-2002 results*" and this dataset includes Fischer's [41] "*the non-Cu area*" data.

In Figure 5, we fitted this largest publicly available dataset for Nb$_3$Sn conductors fabricated using bronze technology to Equation (15). The deduced parameters were $|F_{p,max}(0)| = 74 \pm 3 \frac{GN}{m^3}$, and $\delta = 176 \pm 12$ nm. The parameters have low dependence (~ 0.87), which indicates that our model (Equation (15)) is not over-parameterized.

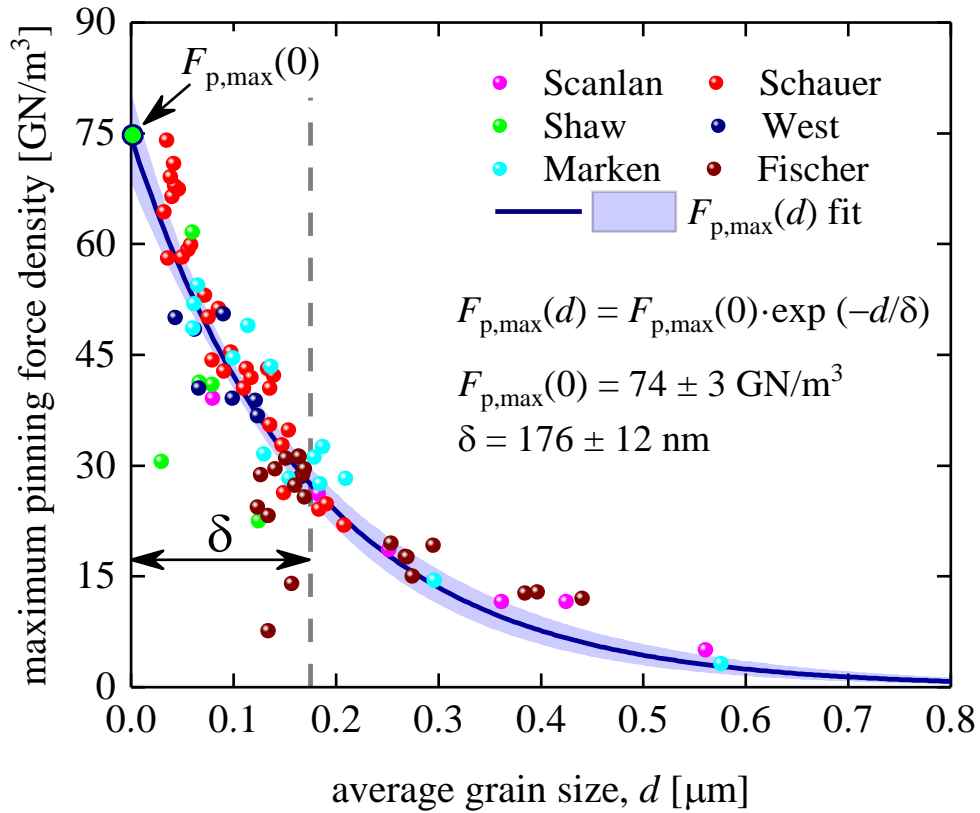

**Figure 5.** Maximum pinning force density, $|F_{p,max}(d)|$, vs average grain size, $d$, for the non-Cu Nb$_3$Sn wires and data fit to Equation (15). Raw data reported by Marken [42], West *et al* [43], Fischer [40], Shaw

[44], Schauer et al [45], and Scanlan et al [46]. Nb$_3$Sn conductors were fabricated by bronze technology. Deduced parameters are $|F_{p,max}(0)| = 74 \pm 3 \frac{GN}{m^3}$, $\delta = 176 \pm 12$ nm; fit quality is 0.9248. 95% confidence bands are shown by grey shadow areas.

### 3.2. Powder-in-tube technology samples

Powder-in-tube technology for Nb$_3$Sn-based wires has been described in detail elsewhere [1]. For our analysis, we used $|F_{p,max}(d)|$ dataset reported by Fischer [40] and Xu et al [57]. In Figure 6, we show the results of the fit of this dataset to Equation (15).

It is interesting to note that the deduced $\delta = 175 \pm 13$ nm is in remarkable agreement with its counterpart deduced for samples fabricated by bronze technology. The deduced parameters also have low dependence (~ 0.87), which is an additional indication that our model (Equation (15)) is not over-parameterized.

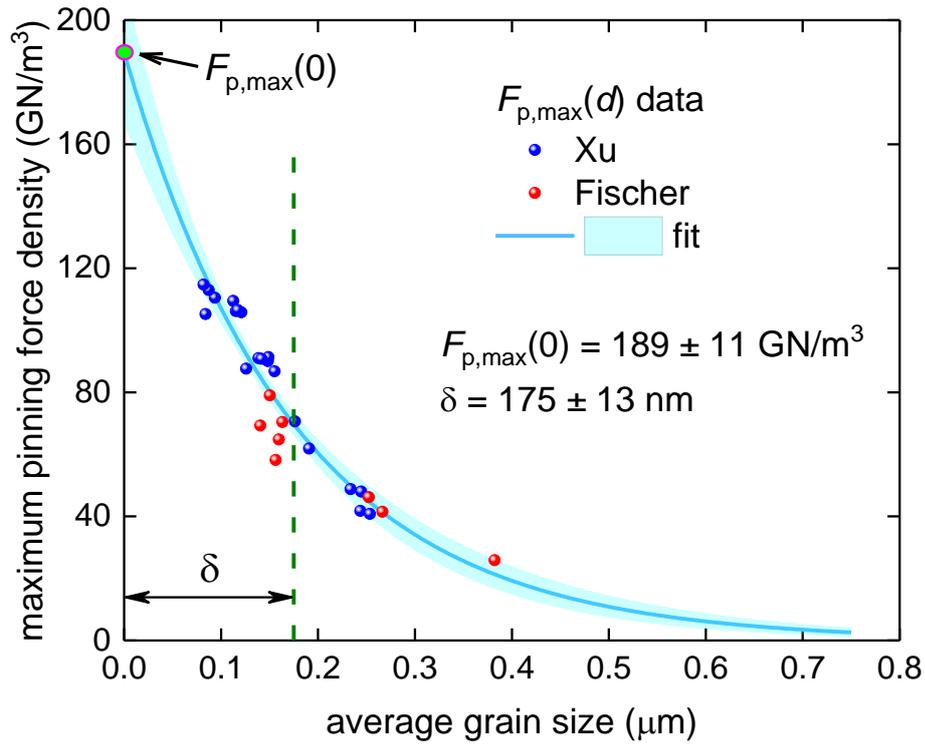

**Figure 6.** Maximum pinning force density, $|F_{p,max}(d)|$, vs average grain size, $d$, for the A15 layer fabricated by powder-in-tube technology [40,57]. Raw data reported by Fischer [40] and Xu et al [57]. Deduced parameters are $|F_{p,max}(0)| = 189 \pm 11 \frac{GN}{m^3}$, $\delta = 175 \pm 13$ nm; fit quality is 0.9093. 95% confidence bands are shown by pink shadow areas.

### 3.3. Samples fabricated by Flükiger et al by bronze technology [38]

Flükiger *et al* [38] reported full $|\vec{F_p}(B)|$ curves, which we analysed in Figure 1, for four samples fabricated using bronze technology. It should be noted that this research group utilized a different normalization procedure for the absolute value of the pinning force density from that used by other research groups [40,42–46]. Therefore, we analyzed this dataset separately (Figure 7). Although this dataset has only four $|F_{p,max}(d)|$ data points, we fitted this dataset to Equation (15) to estimate the robustness of our approach for extracting the characteristic length, $\delta$, from limited $|F_{p,max}(d)|$ datasets. The deduced $\delta = 146 \pm 15$ nm is in the same ballpark as the $\delta$ values deduced from the fits to Equation (15) for large datasets (Figures 5, 6).

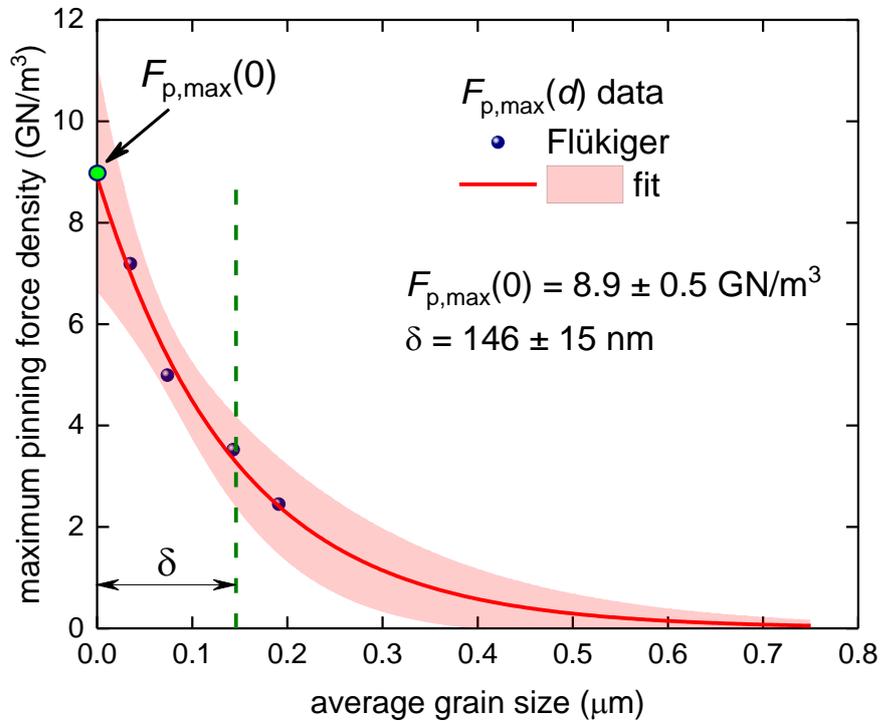

**Figure 7.** Maximum pinning force density, $|F_{p,max}(d)|$, vs average grain size, $d$, for samples fabricated by bronze technology and data fit to Equation (15). Raw data reported by Flükiger *et al* [38]. Deduced parameters are $|F_{p,max}(0)| = 8.9 \pm 0.5 \frac{\text{GN}}{\text{m}^3}$, $\delta = 146 \pm 15$ nm. fit quality is 0.9837. 95% confidence bands are shown by pink shadow areas.

### 4. Discussion

Primary result of our analysis is that Nb$_3$Sn conductors exhibit fundamental length constant, $\delta$, which is in the range of 146 nm $\leq \delta \leq$ 176 nm, and which characterizes maximal intrinsic in-field performance of real world multifilamentary Nb$_3$Sn-based wires.

Our current understanding of this unexpected result can be explained by two hypotheses, both of which are based on the interpretation that one of the two multiplication terms in the formal definition of the pinning force density (Equation (1)), $\vec{F_p}(J_c, B) = \vec{J_c} \otimes \vec{B}$, exhibits exponential decay with characteristic length $\delta$. Thus, there are two possible scenarios/mechanisms:

### 4.1. Exponential dependence of the $|\vec{J_c}|$ vs grain size at $|F_{p,max}|$

This interpretation is based on an analog to the exponential decay $\sim e^{-\frac{x}{\lambda}}$ (more accurately $\sim \frac{\cosh\left(\frac{x}{\lambda}\right)}{\cosh\left(\frac{d}{\lambda}\right)}$ dependence, where $d$ is the slab half-thickness and the layer thickness $\lambda$ is London penetration depth [55]) of the self-field transport current density from the superconductor-vacuum interface, which is the London's law. Considering that under high-field conditions, the interfaces in polycrystalline $Nb_3Sn$ are grain boundaries, we naturally came to Equation (15), where the thickness of the layer (where the dissipative-free transport current flows at the condition of the pinning force maximum) is the characteristic length $\delta$.

A schematic representation of $\delta$-layers in the polycrystalline $Nb_3Sn$ phase, where we drew the $\delta$-layer, is shown in Figure 8.

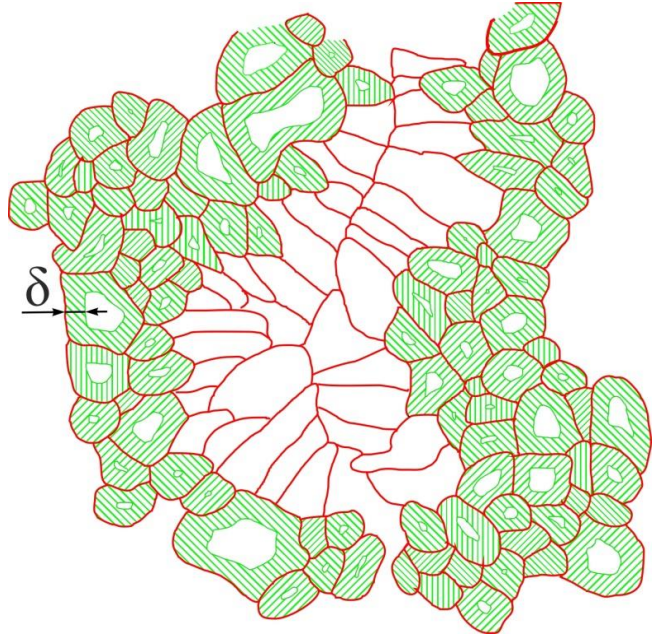

**Figure 8.** Schematic representation of the effective areas ($\delta$-layer) in a cross-section of the eqiaxed $Nb_3Sn$ layer.
16

In this interpretation, large-size grains, $d \gg \delta$, are less effective areas to carry dissipative-free transport current, because central areas of these large grains do not contribute in transferring the transport current (Figure 8), and the current density is reduced by the exponential law. At the same time, small grains, $d \leq \delta$, are very effective areas for carrying dissipative-free transport current flow (Figure 8), because the full grain cross-section area works with approximately the same efficiency.

## 4.2. Exponential dependence of the $|\vec{B}|$ vs grain size at $|F_{p,max}|$

Alternative interpretation is based on an assumption that the flux pinning potential has exponential dependence $\sim e^{-\frac{x}{\delta}}$. As a result, the dissipative-free current can flow only within a thin layer (the thickness of $\delta$) from both sides of grain boundaries, because the flux pinning is strong there and vortices can be hold by the potential vs the Lorentz force. In this interpretation, central areas of large-size grains, $d \gg \delta$, also do not contribute to transfer dissipative-free in-field transport current, because vortices are not hold strong enough vs the Lorentz force. While, the small-size grains, $d \leq \delta$, are very effective to carry dissipative-free transport current flow (Figure 8), because vortices are pinned by pinning potential across full grain area cross-section.

It is interesting to note that the schematic for the effective areas that can carry dissipative-free transport current is the same for both scenarios (Figure 8).

Thus, our current interpretation of the result is that the highest performance of the in-field transport current capacity of Nb$_3$Sn wires is determined by the thin layer with characteristic thickness of $\delta \cong 175$ nm which surrounds the grain boundaries from both sides.

## 5. Refined model

Both our interpretations (Sections 4.1, 4.2, and Figure 8) of the derived characteristic length $\delta$ are based on the idea of exponential spatial decay of the $F_{p,max}$ toward the grain center. Based on this, we came to a need to refine our primary fitting equation (Eq. 15) to account for the fact that the grain boundary surrounds the grain body and the total integrated $F_{p,max}$ associated with the whole grain is a two-dimensional integral from the $F_{p,max}(x,y)$ function within each grain (in this axis arrangement, we assume that the transport current is flowing along the $z$-axis).

In an attempt to compromise between the complexity of mathematics and the accuracy of describing the grain boundary network (observed experimentally [58], Figure 8) we developed a square lattice model (Figure 9). This model can serve as a model on which the basic properties of other more accurate models can be depicted.

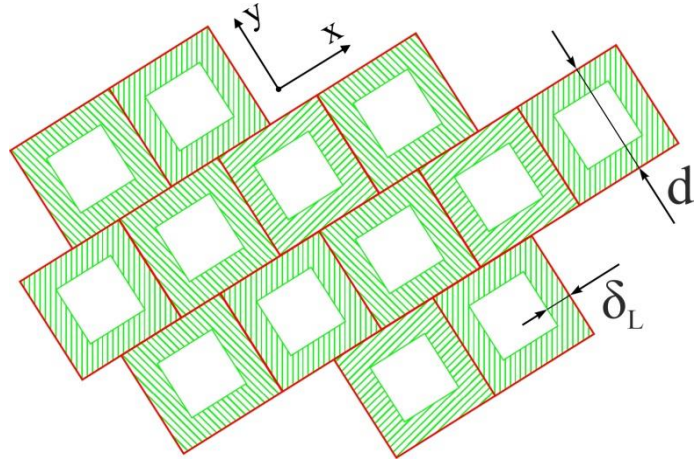

**Figure 9.** Schematic representation of the square grain model, where grain boundaries are indicated in red, and the characteristic thickness is green. The characteristic layer thickness, $\delta_L$, and grain size, $d$, are also shown.

In the square lattice model (Figure 9), the global maximum of the pinning force density for the sample, $F_{p,max,global}$, is the sum of the equal terms, which are the integrated maximum of the pinning force density of each grain:

$$F_{p,max,global}(d) = \sum_{n=1}^{N} \left( \int_{Area} F_{p,max,grain}(x,y,d)dxdy \right)_n = N \times$$

$$\int_{area} F_{p,max,grain}(x,y,d)dxdy = N \times F_{p,max,single\ grain}(d). \qquad (18)$$

In accordance with Eq. 18, the primary task is to calculate the integrated $F_{p,max}$ over the area of one grain as follows:

$$F_{p,max,single\ grain}(d) = \int_{area} F_{p,max,grain\ area}(x, y, d) dx dy. \tag{19}$$

First, consider a one-dimensional problem (for which the grain is within $\left(-\frac{d}{2}, \frac{d}{2}\right)$) for which we can use the London theory solution for exponentially decaying current density or magnetic flux density for rectangular slabs [55,59–61]:

$$F_{p,max,grain\ width}(x, d) = F_{p,max,grain\ boundary} \times \frac{\cosh\left(\frac{x}{\delta}\right)}{\cosh\left(\frac{d}{2\delta}\right)}. \tag{20}$$

The integration of Eq. 20 gives:

$$\int_{-\frac{d}{2}}^{\frac{d}{2}} F_{p,max,grain\ width}(x, d) dx = F_{p,max,grain\ boundary} \times \left(\frac{2\delta}{d}\right) \times \tanh\left(\frac{d}{2\delta}\right). \tag{21}$$

The solution for *y*-axis is similar to that for *x*-axis:

$$\int_{-\frac{d}{2}}^{\frac{d}{2}} F_{p,max,grain\ length}(y, d) dy = F_{p,max,grain\ boundary} \times \left(\frac{2\delta}{d}\right) \times \tanh\left(\frac{d}{2\delta}\right). \tag{22}$$

Considering that the London equation is a linear differential equation, the sum of two partial solutions is also the solution of the equation, and the solution for the two-dimensional problem can be written in the form:

$$F_{p,max,single\ grain}(d) = \int_{-\frac{d}{2}}^{\frac{d}{2}} F_{p,max,grain\ width}(x, d) dx +$$

$$\int_{-\frac{d}{2}}^{\frac{d}{2}} F_{p,max,grain\ length}(y, d) dy = 2 \times F_{p,max,grain\ boundary} \times \left(\frac{2\delta_L}{d}\right) \times \tanh\left(\frac{d}{2\delta_L}\right), \tag{23}$$

where $\delta_L$ is the designation for the characteristic length of the two-dimensional square model (Figure 9), which is based on the London theory solution for current density and magnetic flux density distributions for a square slab [55,59–61].

Thus, the final fitting equation for the refined model is:

$$F_{p,max,global}(d) \equiv F_{p,max}(d) = F_{p,max}(0) \times \left(\frac{2\delta_L}{d}\right) \times \tanh\left(\frac{d}{2\delta_L}\right), \tag{24}$$

where $F_{p,max}(0) = N \times 2 \times F_{p,max,grain\ boundary}$, and the other terms are defined above.

$F_{p,max}(d)$ dataset (which we have already analyzed in Figs. 2,4,5) and data fit to Eq. (24) are shown in Figure 10.

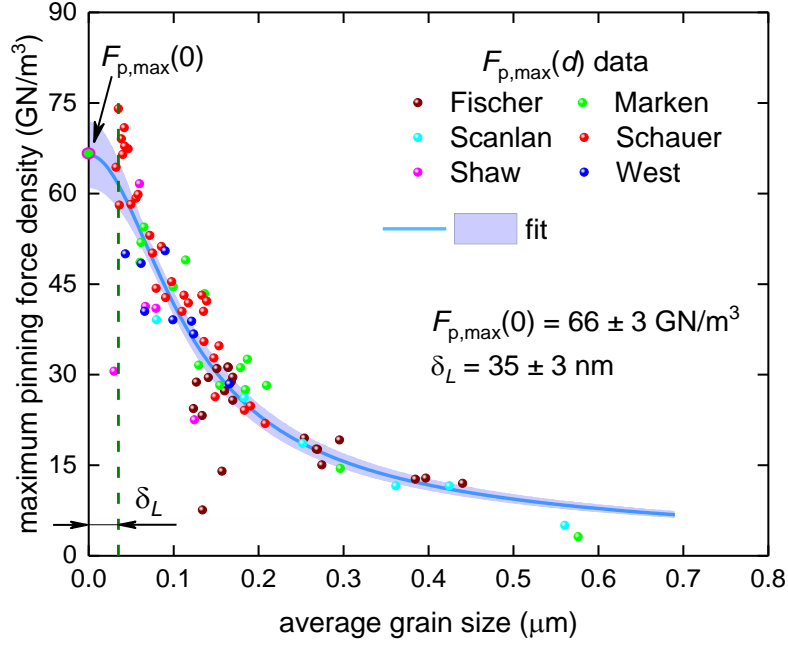

**Figure 10.** Maximum pinning force density, $|F_{p,max}(d)|$, vs. average grain size, $d$, for the non-Cu Nb$_3$Sn wires and the data fit to Equation (24). Raw data reported by Marken [42], West *et al* [43], Fischer [40], Shaw [44], Schauer *et al* [45], and Scanlan *et al* [46]. Nb$_3$Sn conductors were fabricated using bronze technology. Deduced parameters are $|F_{p,max}(0)| = 66 \pm 3 \frac{\text{GN}}{\text{m}^3}$, $\delta_L = 35 \pm 3$ nm; fit quality is 0.8100. 95% confidence bands are shown as dark yellow shadow areas.

$F_{p,max}(d)$ dataset for Nb$_3$Sn powder-in-tube technology samples [40,57] (which we have already analyzed in Figure 6) and data fit to Eq. (24) are shown in Figure 11.

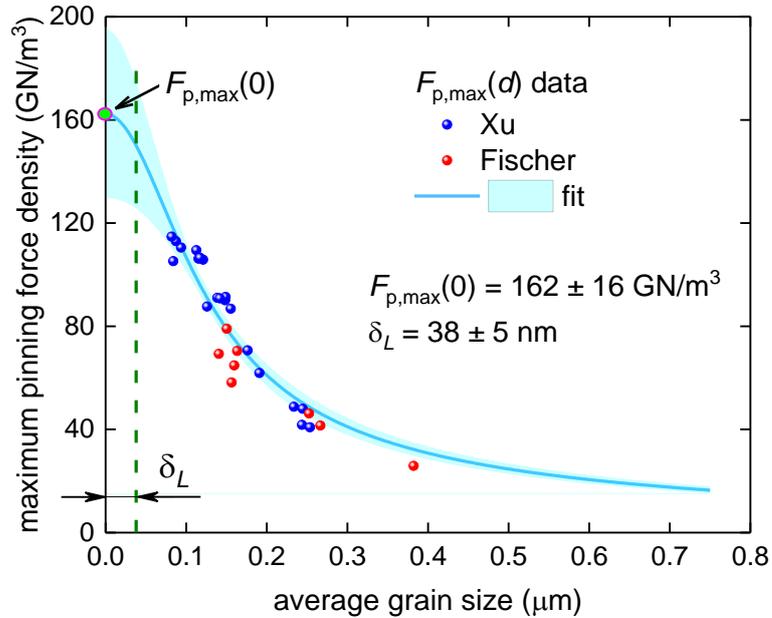

**Figure 11.** Maximum pinning force density, $|F_{p,max}(d)|$, vs. average grain size, $d$, for the A15 layer fabricated by the powder-in-tube technology [40,57]. Raw data reported by Fischer [40] and Xu *et al* [57]. Data fit to Equation (24). Deduced parameters are $|F_{p,max}(0)| = 162 \pm 16 \frac{\text{GN}}{\text{m}^3}$, $\delta = 38 \pm 5$ nm; fit quality is 0.8966. 95% confidence bands are shown by pink shadow areas.

$F_{p,max}(d)$ dataset for samples reported by Flükiger *et al* [38] (which we have already analyzed in Figure 7) and data fit to Eq. (24) are shown in Figure 12.

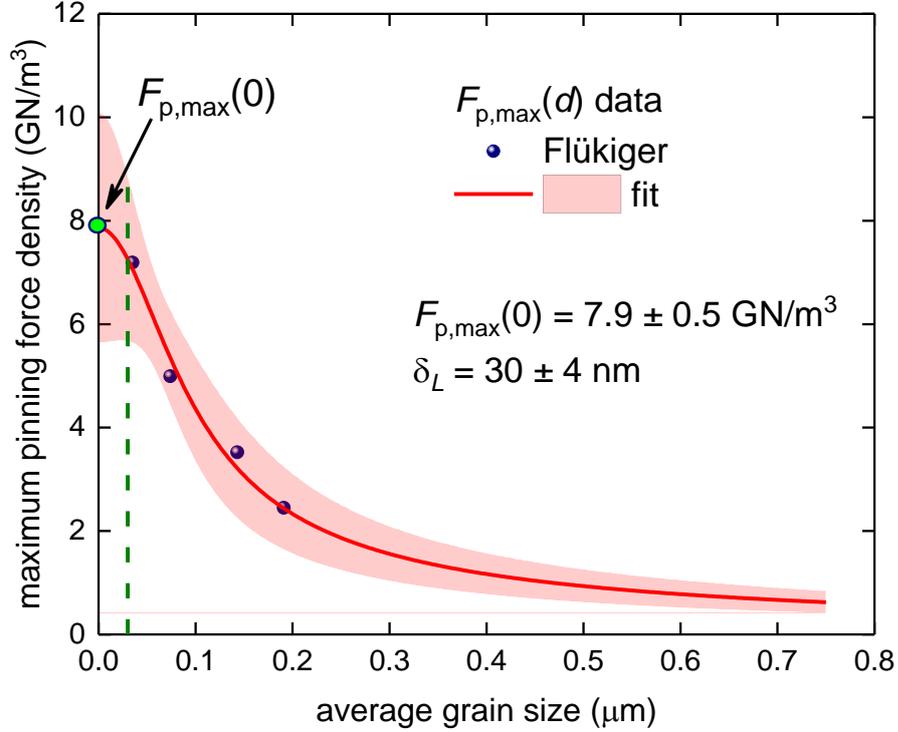

**Figure 12.** The maximum pinning force density, $|F_{p,max}(d)|$, vs average grain size, $d$, for samples fabricated using bronze technology and data fit to Equation (24). Raw data reported by Flükiger *et al* [38]. Deduced parameters are $|F_{p,max}(0)| = 7.9 \pm 0.5 \frac{GN}{m^3}$, $\delta_L = 30 \pm 4$ nm; fit quality is 0.9812. 95% confidence bands are shown by pink shadow areas.

The derived $\delta_L$ from the three available $F_{p,max}(d)$ datasets (Figs. 10-12) are within a narrow range:

$$30 \pm 4 \text{ nm} \leq \delta_L(T = 4.2 \text{ K}) \leq 38 \pm 5 \text{ nm}, \qquad (25)$$

which is the evidence that $\delta_L$ perhaps represents a universal constant for the Nb$_3$Sn phase, which is independent of the sample manufacturing technique.

If so, we can make a comparison of the $\delta_L(T = 4.2\ K)$ with two other fundamental lengths of any superconductors, i.e. the London penetration depth, $\lambda(T = 4.2\ K)$, and the coherence length, $\xi(T = 4.2\ K)$:

$$\frac{\delta_L(T=4.2\ K)}{\lambda(T=4.2\ K)} = \frac{\frac{\delta_L(T=4.2\ K)}{\lambda(0)}}{\sqrt{1-\left(\frac{T=4.2\ K}{T_C=18\ K}\right)^4}} = 0.53 \cong \frac{1}{2}, \qquad (26)$$

$$\frac{\delta_L(T=4.2\ K)}{\xi(T=4.2\ K)} = \frac{\frac{\delta_L(T=4.2\ K)}{\xi(0)}}{\sqrt{1-\left(\frac{T=4.2\ K}{T_C=18\ K}\right)}} = 9.9 \cong 10, \qquad (27)$$

where we used the two-fluid (Chapter 2 in [55]), and the Ginzburg-Landau theory expressions (Chapter 6 in [55]), and $\lambda(0) = 65\ nm$ (Table 12.1, p. 343 in [55]), $\xi(0) = 3\ nm$ (Table 12.1, p. 343 in [55]), $T_c = 18\ K$ (Table 12.1, p. 343 in [55]), and average value for the derived characteristic length $\overline{\delta_L}(T = 4.2\ K) = 34\ nm$ (Figures 10-12).

The obtained ratios (Equations 26, 27) demonstrate that the derived $\delta_L$ is half the London penetration depth, $\lambda(T)$, and one order of magnitude larger than the coherence length, $\xi(T)$.

This implies that $\delta_L$ might represent a new characteristic length constant for superconductors, which lies between $\lambda(T)$ and $\xi(T)$ for high-$\kappa$ superconductors and is associated with the maximum superconductor performance in an applied magnetic field.

## 6. Conclusions

Finning force maximum, $|F_{p,max}(J_c, B)|$, represents global maximum of the vector product of the transport critical current density, $\vec{J_c}$, and the applied magnetic field, $\vec{B}$, and it can be derived as from $|F_p(B)|$ [35–37], as from $|F_p(J_c)|$ [62] projections of the $F_p(J_c, B)$ function (Equation (1)).

In this report we re-analysed experimental data on the dependence of the maximum pinning force density, $|F_{p,max}(d, T = 4.2\ \text{K})|$ (deduced from the $|F_p(B)|$ [35–37] projection) from the average grain size in practical low-$T_c$ multifilamentary Nb$_3$Sn conductors [1–34,38–46,55,56,62] fabricated by bronze and power-in-tube technologies.

The primary result of our analysis is that Nb$_3$Sn conductors at their maximum in-field performance exhibit characteristic length $\delta = 175$ nm, which is the same for samples fabricated by bronze and powder-in-tube technologies.

Following an interpretation that there is a characteristic thickness of the layer surrounding the grain boundary network, where a dissipative-free transport current flows, we developed London-like model in assumption of square $Nb_3Sn$ grains, for which the characteristic length, $\delta_L$, was deduced in the range:

$$30 \pm 4 \text{ nm} \leq \delta_L(T = 4.2 \text{ K}) \leq 38 \pm 5 \text{ nm}. \tag{28}$$

The comparison of the derived $\delta_L$ with two fundamental characteristic lengths of $Nb_3Sn$, i.e. the London penetration depth, $\lambda$, and the coherence length, $\xi$, shows that:

$$\frac{\delta_L}{\lambda} = \frac{1}{2}, \tag{30}$$

$$\frac{\delta_L}{\xi} = 10, \tag{31}$$

which is more likely implied that $\delta_L$ represents a new characteristic length constant for superconductors, which lies between $\lambda(T)$ and $\xi(T)$ for high-$\kappa$ superconductors, and which is associated with the maximum superconductor performance in applied magnetic field.

**Author Contributions:** E.F.T. conceived the work and proposed exponential and hyperbolic tangent dependence for $F_{p,max}(d)$, E.F.T. and E.G.V.-Z. searched publicly available experimental data and performed data fit and calculations, E.F.T. proposed to interpret $\delta$ as the characteristic thickness for transport current flow, E.G.V.-Z. proposed to interpret $\delta$ as the characteristic length for flux pinning potential. All authors discussed results. E.G.V.-Z. prepared final figures. E.F.T. wrote the manuscript, which was revised by E.G.V.-Z., I.L.D. and E.N.P.

**Funding:** The research was carried out within the state assignment of Ministry of Science and Higher Education of the Russian Federation (theme "Pressure" No. 122021000032-5). E.F.T. thanks the research funding from the Ministry of Science and Higher Education of the Russian Federation (Ural Federal University Program of Development within the Priority-2030 Program).

**Conflicts of Interest:** The authors declare no conflict of interest. The funders had no role in the design of the study; in the collection, analyses, or interpretation of data; in the writing of the manuscript; or in the decision to publish the results.